\documentclass[a4paper,11pt,reqno]{article}
\usepackage{a4wide}
\def\cA{\mathcal{A}}
\def\pzero{{\cal F}_{(0)}}
\def\pone{{\cal F}_{(1)}}
\def\ptwo{{\cal F}_{(2)}}
\def\cB{{\cal{B}}}

\def\cD{{\cal D}}

\def\arctanh{\rm arctanh}

\usepackage{amsmath,amsfonts,amssymb}
\usepackage[T1]{fontenc}
\usepackage[p,osf]{cochineal}
\usepackage[varqu,varl,var0]{inconsolata}
\usepackage[scale=.95,type1]{cabin}
\usepackage[vvarbb]{newtxmath}
\usepackage[cal=boondoxo]{mathalfa}
\usepackage[mathscr]{euscript}
\usepackage{bbold} 

\usepackage{pict2e}

\makeatletter
\DeclareRobustCommand{\loplus}{\mathbin{\mathpalette\dog@lsemi{+}}}
\DeclareRobustCommand{\lotimes}{\mathbin{\mathpalette\dog@lsemi{\times}}}
\DeclareRobustCommand{\roplus}{\mathbin{\mathpalette\dog@rsemi{+}}}
\DeclareRobustCommand{\rotimes}{\mathbin{\mathpalette\dog@rsemi{\times}}}

\newcommand{\dog@rsemi}[2]{\dog@semi{#1}{#2}{-90,90}}
\newcommand{\dog@lsemi}[2]{\dog@semi{#1}{#2}{270,90}}
\newcommand{\dog@semi}[3]{%
  \begingroup
  \sbox\z@{$\m@th#1#2$}%
  \setlength{\unitlength}{\dimexpr\ht\z@+\dp\z@\relax}%
  \makebox[\wd\z@]{\raisebox{-\dp\z@}{%
    \begin{picture}(1,1)
    \linethickness{\variable@rule{#1}}
    \roundcap
    \put(0.5,0.5){\makebox(0,0){\raisebox{\dp\z@}{$\m@th#1#2$}}}
    \put(0.5,0.5){\arc[#3]{0.5}}
    \end{picture}%
  }}%
  \endgroup
}
\newcommand{\variable@rule}[1]{%
  \fontdimen8  
  \ifx#1\displaystyle\textfont3\else
    \ifx#1\textstyle\textfont3\else
      \ifx#1\scriptstyle\scriptfont3\else
        \scriptscriptfont3\relax
  \fi\fi\fi
}
\makeatother

\usepackage{lettrine}

\usepackage{nicefrac}
\usepackage{setspace}
\usepackage{datetime}
\usepackage[nosort]{cite}
\usepackage{upgreek}

\usepackage[euler]{textgreek}
\usepackage[breaklinks=true]{hyperref}
\hypersetup{
			colorlinks=true,
			urlcolor=magenta,
			citecolor=magenta,
			linkcolor=magenta,
     }

\let\oldsqrt\sqrt
\def\sqrt{\mathpalette\DHLhksqrt}
\def\DHLhksqrt#1#2{%
\setbox0=\hbox{$#1\oldsqrt{#2\,}$}\dimen0=\ht0
\advance\dimen0-0.2\ht0
\setbox2=\hbox{\vrule height\ht0 depth -\dimen0}%
{\box0\lower0.4pt\box2}}

\newcommand{\RNum}[1]{\uppercase\expandafter{\romannumeral #1\relax}}
\usepackage{xcolor}
\usepackage{changes}
\reversemarginpar
\definechangesauthor[name=DRB, color=purple]{DRB}
\definechangesauthor[name=GAH, color=cyan]{GAH}
\definechangesauthor[name=FD, color=olive]{FD}
\author{
  \begin{minipage}{.99\linewidth}
    \vspace{1cm}
       \begin{center}
      \begin{small}
      \textbf{Gabriel Arenas--Henriquez$^1$\footnote{arenas2023@tsinghua.edu.cn}\, , Felipe Diaz$^2$\footnote{fdiaz47@gmail.com}\,\,~, and 
      \textbf{David Rivera--Betancour}}$^{2}$\footnote{driverabetancour@itmp.msu.ru}
              \end{small}
    \end{center}
    \vspace{0.5cm}
    \hspace{1cm}\begin{minipage}{.55\linewidth}
\begin{center}     {\it \begin{footnotesize}
\hbox{\kern-0.1cm\vbox{\vskip0cm
 \begin{itemize}
  \item[$^1$]  Yau Mathematical Sciences Center\\ Tsinghua University \\ Beijing 100084, China \\                     
                           \vskip0.25cm
      \end{itemize}}
\kern-1.5cm\vbox{
\begin{itemize}
  \item[$^2$]Institute for Theoretical and Mathematical Physics, \\
Moscow State University, 119991,\\ Leninskie Gory, GSP-1, Moscow, Russia
      \end{itemize}
      \vskip0.cm
}}
     \end{footnotesize}}
\end{center}
    \end{minipage}
    \vspace{0.5cm}\begin{minipage}{.7\linewidth}
     \end{minipage}
  \end{minipage}
}

\title{\vspace{2.5cm}
 \boldmath \begin{LARGE}
    \textbf{\textsc{Generalized Fefferman—Graham gauge \\ and boundary Weyl structures}}
  \end{LARGE} \unboldmath
}

\date{}

\begin{document}

%\blinddocument
%\blindmathpaper

\begin{titlepage}
\maketitle
\thispagestyle{empty}

 %\vspace{-12.cm}
  %\begin{flushright}
  %CPHT-RR022.042022\\
  %\end{flushright}
 %\vspace{11.cm}

\begin{center}
\textsc{Abstract}\\  
\vspace{1. cm}	
\begin{minipage}{1.0\linewidth}
{In the framework of AdS/CFT correspondence, the Fefferman--Graham (FG) gauge offers a useful way to express asymptotically anti-de Sitter spaces, allowing a clear identification of their boundary structure. A known feature of this approach is that choosing a particular conformal representative for the boundary metric breaks explicitly the boundary scaling symmetry. Recent developments have shown that it is possible to generalize the FG gauge to restore boundary Weyl invariance by adopting the Weyl--Fefferman--Graham gauge. In this paper, we focus on three-dimensional gravity and study the emergence of a boundary Weyl structure when considering the most general AdS boundary conditions introduced by Grumiller and Riegler \cite{Grumiller:2016pqb}. We extend the holographic renormalization scheme to incorporate Weyl covariant quantities, identifying new subleading divergences appearing at the boundary. To address these, we introduce a new codimension-two counterterm, or corner term, that ensures the finiteness of the gravitational action. From here, we construct the quantum-generating functional, the holographic stress tensor, and compute the corresponding Weyl anomaly, showing that the latter is now expressed in a full Weyl covariant way. Finally, we discuss explicit applications to holographic integrable models and accelerating black holes. For the latter, we show that the new corner term plays a crucial role in the computation of the Euclidean on-shell action.}

\end{minipage}
\end{center}

%\vspace{2cm} 

\end{titlepage}

\onehalfspace
%\noindent\rule{\textwidth}{1.2pt}
%\vspace{-1cm}
\begingroup
\hypersetup{linkcolor=black}
\tableofcontents
\endgroup
\noindent\rule{\textwidth}{0.6pt}

%%
%%
%%

%\noindent\rule{\textwidth}{1.2pt}
%\vspace{-1cm}
\begingroup
\hypersetup{linkcolor=black}

\endgroup
%\noindent\rule{\textwidth}{0.6pt}
%%%%%%%%%%%%%%%%%%%%%%%%%%%%%%%%
%%%%%%%%%%% INTRO %%%%%%%%%%%%%%
%%%%%%%%%%%%%%%%%%%%%%%%%%%%%%%%
\section{Introduction}\label{Sec:Intro}
%%%%%%%%%%%
%%%%%%%%%%%
Gauge/Gravity duality \cite{Maldacena:1997re, Gubser:1998bc, Witten:1998qj} has emerged as a powerful tool in theoretical physics that provides profound insights into strongly correlated systems by relating them to classical gravity theories.
Initially formulated as a correspondence between type IIB string theory on AdS$_{5} \times S^{5}$ and $\cal{N}=4$ super Yang--Mills theory, it has been extended to a wide variety of contexts ranging from high-energy physics to condensed matter systems, and even to quantum information theory. 
One particularly fruitful application of this duality is the fluid/gravity correspondence \cite{Bhattacharyya:2007vjd,Haack:2008cp,Rangamani:2009xk,Hubeny:2011hd}, which connects the dynamics of long-wavelength perturbations in AdS space to the one of a relativistic conformal fluid defined at the timelike boundary. 
By projecting Einstein's equations onto the boundary, the bulk metric can be expressed in a derivative expansion, where the boundary stress tensor emerges as a gradient expansion in terms of hydrodynamic variables. 
This correspondence not only provides a geometric interpretation of fluid dynamics but also offers a systematic way to compute transport coefficients, such as shear viscosity, in strongly coupled systems. In particular, the conformal structure of these field theories naturally incorporates Weyl invariance which imposes constraints on the structure of the fluid stress tensor and its associated transport coefficients. As such, a proper holographic description of these systems must incorporate Weyl covariance as a central feature.

The prominent role of boundary Weyl covariance in the reconstruction of the spacetime is not only restricted to gravity with negative cosmological constant. In recent years, it has been proposed that the dual of an asymptotically (locally) flat spacetime corresponds to a Carrollian conformal field theory (CCFT) defined at null infinity. The latter emerges from the vanishing speed of light limit of a relativistic conformal field theory \cite{Duval:2014uva,Donnay:2022aba,Donnay:2022wvx,Alday:2024yyj}. It has been shown in \cite{Ciambelli:2018wre,Campoleoni:2023fug}, that any Ricci-flat spacetime can be reconstructed order by order in a radial expansion in terms of Carrollian boundary data. This was well implemented in the covariant Newman--Unti gauge, which is obtained through a relaxation of the Newman gauge that makes the covariance with respect to the boundary explicit (Carroll diffeomorphisms and Weyl scalings). In this way, the spacetime is reconstructed in terms of the boundary Carroll geometry, the Carollian momenta of the boundary theory, and an infinite tower of arbitrary Carroll and Weyl covariant tensors. In this procedure, Weyl covariance provides a guiding principle to organize the Carrollian data appearing at each order in the asymptotic expansion.

In the standard holographic framework describing asymptotically AdS (AAdS) spaces, the Weyl symmetry is encoded naturally within the FG gauge. Here, the boundary of AdS is represented by a conformal class of boundary metrics $[\gamma^{(0)}_{ij}]$ which serves as source of the boundary field theory stress tensor \cite{Henningson:1998gx,Balasubramanian:1999re,deHaro:2000vlm}. 
In practice, however, computing observables in the boundary field theory often requires making a specific choice of representative of the boundary metric (or boundary conditions), explicitly breaking the scaling symmetry. To address this limitation, the Weyl--Fefferman--Graham (WFG) gauge was recently introduced \cite{Ciambelli:2019bzz,Jia:2021hgy,Jia:2023gmk,Ciambelli:2023ott,Parisini:2023nbd}. This gauge choice extends the FG gauge allowing cross-terms in the line element between radial and boundary coordinates, thereby introducing a boundary Weyl connection. The boundary theory then, is not only described by its metric but also by its Weyl connection that generates a Weyl current. In this way, the WFG gauge provides a formalism that preserves Weyl covariance at the boundary, allowing a more natural description of systems where local scale invariance plays an important role. 

In three dimensions, gravity is classically trivial as it lacks propagating degrees of freedom. Nevertheless, it serves as an important toy model of quantum gravity, being describable as a gauge theory \cite{Witten:2007kt}. Considering the negative cosmological constant, the problem of finding the associated Hilbert space is mapped to finding the full space of compatible conformal blocks of a two-dimensional holomorphically factorized CFT. Nonetheless, Maloney and Witten \cite{Maloney:2007ud}, using conformal bootstrap techniques, have shown that the sum of known contributions to the partition function from classical geometries can be computed exactly finding an infinite sum of negative norm states making the theory non-physical. The sum is constructed by analyzing the classification of hyperbolic three-manifolds and considering excited states coming from Virasoro descendants that are compatible with modular invariance of the partition function (parametrized by the left and right central charges) following the Brown and Henneaux arguments \cite{Brown:1986nw}. Then,  one needs to include new contributions such as topological defects, contributing with codimension-two terms in the Euclidean action and corresponding to massive boundary states, or other black hole states beyond the BTZ and Brown--Henneaux boundary excitations. 
As three-dimensional gravity is topological, all vacuum solutions are isomorphic to global AdS. Nonetheless, the map between the FG gauge and the WFG gauge is a charged diffeomorphism \cite{Ciambelli:2023ott} showing that both can be identified as physically inequivalent configurations, giving new contributions to the dual theory and possibly to the Maloney--Witten sum. Then, relaxing AdS boundary conditions opens the possibility of having a more complete holographic dictionary with new observables. Indeed, considering the WFG gauge the asymptotic symmetry algebra now contains new contributions coming from charges associated with the Weyl connection generating an extra Heisenberg algebra, with a time-dependent central charge, in direct sum with the usual two copies of the Virasoro algebra (see also \cite{Alessio:2020ioh}) just as in the Bondi--Weyl gauge \cite{Geiller:2021vpg}. 
Interestingly, the falloff of the WFG gauge can be relaxed leading to  -- an older -- choice of boundary conditions \cite{Grumiller:2016pqb}. In this context, the gauge was referred to as generalized Fefferman--Graham (gFG) and corresponds to the most general set of boundary conditions in three-dimensional AdS gravity, being able to reduce to previously known cases such as \cite{Brown:1986nw,Compere:2013bya,Afshar:2016wfy,Troessaert:2013fma,Avery:2013dja}. 

In this work, we close the bridge between these two falloff conditions by showing that the gFG gauge also gives rise to a boundary Weyl structure. We perform holographic renormalization techniques to analyze the asymptotic behaviour of the on-shell gravitational action, identifying new subleading divergences that appear at the boundary. Consequently, we construct the appropriate boundary counterterm that ensures the finiteness of the action. Employing the renormalized action, we obtain the holographic stress tensor expressing it in terms of Weyl covariant objects in which the boundary Weyl current plays a relevant role. 
This machinery will allow us to explore asymptotically non-trivial spacetimes such as accelerating black holes, which are not naturally aligned with the FG gauge. 

This paper is organized as follows: In \autoref{Sec:WFG}, we review essential aspects of the FG gauge, comparing it with the WFG gauge, and briefly illustrate the emergence of Weyl geometry on the AdS boundary and its role in holography. In \autoref{Sec:gFG}, we introduce the gFG gauge, conduct an asymptotic analysis of Einstein’s equations, and examine the on-shell action, where we propose a novel counterterm to regularize additional subleading divergences resulting from cross terms between radial and boundary coordinates in the metric. We derive the holographic stress tensor and Weyl current, emphasizing their contributions to the trace anomaly and central charge in the boundary field theory. In \autoref{Sec:Examples}, we explore potential applications to topologically non-trivial spacetimes, such as accelerating black holes, highlighting the broader significance of the gFG gauge in holographic contexts. Finally, in \autoref{Sec:Discussions}, we conclude with open questions and potential future directions. 

%%%%%%%%%%%%%%%%%%%%%%%%%%%%%%%%
%%%%%%%%%%% Weyl %%%%%%%%%%%%%%%
%%%%%%%%%%%%%%%%%%%%%%%%%%%%%%%%
\section{Holography and Weyl geometries}\label{Sec:WFG}

The FG theorem \cite{Fefferman:1984asd, Skenderis:2002wp} states that the line element of a $d+1$-dimensional asymptotically locally AdS (AlAdS) spacetime can be always written as 
\begin{align}\label{FGgauge}
    ds^2 = \frac{\ell^2}{z^2}\left(dz^2 + \gamma_{ij}(z,x)dx^i dx^j\right)~,
\end{align}
where $z$ is usually referred to as the holographic or radial coordinate. The metric has a second-order pole at $z=0$ defining a boundary that is not endowed with a metric structure but rather a conformal class of them, i.e. a conformal boundary \cite{Penrose:1985bww}. 

Near the conformal boundary, the induced metric at constant $z$ admits a power series expansion
\begin{align}
    \gamma_{ij} = \gamma_{ij}^{(0)} + z \gamma_{ij}^{(1)} + z^2 \gamma_{ij}^{(2)} +\ldots + z^d\left(\gamma_{ij}^{(d)} + g_{ij}^{(d)}\log z \right) + \ldots~,
\end{align}
where the $\gamma^{(0)}_{ij}$ coefficient corresponds to the fixed conformal representative of the conformal classes of boundary metrics that source the holographic stress tensor. The logarithmic contribution only appears when the dimension of the boundary, $d$~, is even, and its coefficient $g^{(d)}$ is known as  \textit{obstruction tensor} \cite{Fefferman:1984asd, graham2004ambient} which is traceless symmetric and Weyl covariant \footnote{See \cite{Anastasiou:2020zwc} for a different approach.}. When $d=2$ the obstruction tensor vanishes exactly so there is no need to add such contribution to the expansion, see for instance \cite{Jia:2023gmk} for details. 

Bulk diffeomorphisms naturally induce boundary Weyl transformations. However, the FG gauge \eqref{FGgauge} breaks the boundary Weyl diffeomorphisms by fixing a specific conformal representative $\gamma_{(0)}$~. A particular subgroup of diffeomorphisms, known as Penrose--Brown--Henneaux (PBH) transformation \cite{Penrose:1985bww, Brown:1986nw}, of the form
\begin{align} \label{PBH}
    z\to z' = z/{\cal B}(x)~,\qquad x^i \to x'^i=x^i+b^i(z,x)~
\end{align}
 preserves the form of the FG gauge \eqref{FGgauge} provided that\footnote{Here we consider infinitesimal transformations and work at first order in $\sigma$ and $b^i$~.} \cite{Imbimbo:1999bj}
\begin{align}
    \partial_z b^i = \frac{\ell^2}{2}\gamma^{ij}\partial_j \sigma~,
\end{align}
where $\sigma = \frac12 \log {\cal B}$~. The PBH transformation induces a Weyl transformation on the boundary but breaks the Weyl covariance of the subleading terms of the expansion. To restore Weyl covariance, \cite{Ciambelli:2019bzz} proposes the following modified gauge
\begin{align}
    ds^2 = \ell^2\left(\frac{dz}{z} - a_i dx^i\right)^2 + h_{ij}dx^i dx^j~,
\end{align}
referred to as the Weyl--Fefferman--Graham (WFG) gauge. In this gauge, the line element remains stable under the following Weyl diffeomorphism
\begin{align}\label{WeylDiff}
    z\to z' = z/{\cB(x)}~,\qquad  x^i\to x'^i=x^i~.
\end{align}
Note that there is no need to introduce a compensating boundary transformation, such as the one given by the vector $b^i$ in the PBH case \eqref{PBH}. 
Similarly, as in the FG gauge, we can consider a near boundary expansion of the induced metric $h_{ij}$ and the vector $a_{i}$ 
\begin{align} \label{expansionWFG}
   h_{ij}  = \left(\frac{\ell}{z}\right)^2 \sum_{k\geq 0}h_{ij}^{(k)}~, \qquad a_i = \sum_{k\geq 0} \left(\frac{z}{\ell}\right)^{k}a_i^{(k)}~.
\end{align}
Using the field equations, one can check that all coefficients $h_{ij}^{(k)}$ and $a_i^{k}$ with $k$ odd, identically vanishes.
Furthermore, considering the transformation \eqref{WeylDiff}, we observe that each subleading coefficient in \eqref{expansionWFG} transform covariantly under Weyl rescaling, i.e., 
\begin{align}
    h_{ij}^{(2k)} \to \cB^{2k-2}h_{ij}^{(2k)}~,\qquad a_i\to \cB^{2k}a_i^{(2k)}-\delta_{k,0}\partial_i \log \cB~.
\end{align}
Thus, the terms $k \neq 0$ correspond to boundary Weyl tensors with definite Weyl weights.  Interestingly, the leading term of $a_{i}$, $a_{i}^{(0)}$, undergoes an inhomogeneous transformation resembling that of a Weyl connection. In fact, this coefficient can be identified with the boundary Weyl connection that generates a Weyl current in the holographic theory \cite{Ciambelli:2019bzz}.
To verify this, one can compute the Christoffel symbol of the bulk geometry and expand around $z = 0$, finding that to leading order,
\begin{align}\label{WChris}
\mathring\Gamma_{(0)}^m{}_{ij}:= \frac12 h_{(0)}^{mn}\left(\partial_i h^{(0)}_{nj} + \partial_j h^{(0)}_{ni} - \partial_n h^{(0)}_{ij}\right) - \left(a_i^{(0)}\delta^m_j + a_j^{(0)}\delta^m_i - a_n^{(0)}h_{(0)}^{mn}h_{ij}^{(0)}\right)~.
\end{align}
As can be seen, these coefficients differ from the standard Christoffel symbols as now contain contributions from the Weyl connection $a^{(0)}_i$. Moreover, the coefficients are Weyl invariant and define the unique torsion-free connection\footnote{More generally, one could relax this condition to include a non-vanishing torsion breaking the Weyl manifold condition. This has been recently explored in other holographic contexts, see for instance \cite{Adami:2024rkr}.} with Weyl metricity, viz.
\begin{align}\label{Wmetricity}
    \mathring\nabla^{(0)}_k h_{ij}^{(0)} - 2a_k^{(0)}h_{ij}^{(0)} = 0~,
\end{align}
where $\mathring\nabla^{(0)}$ is the covariant derivative associated with \eqref{WChris}. Although, from the bulk perspective, $a_{i}^{(0)}$ is introduced as a pure gauge artefact, it effectively defines a boundary Weyl connection. 

For a given boundary tensor $P$ (suppressing indices), which transforms under Weyl rescaling as $P\to {\cal B}^w P$, we define its Weyl weight as $w$.
Thus, we define a boundary Weyl covariant derivative as
\begin{align}\label{Weylring}
    \hat\nabla^{(0)}_i P := \mathring\nabla_i^{(0)} P + w a_i^{(0)}P~.
\end{align}
Under Weyl rescalings, the Weyl covariant derivative transforms as 
\begin{align}
    \mathring\nabla_i^{(0)} P + w a_i^{(0)}P \to {\cal B}^w\left(\mathring\nabla_i^{(0)} P + w a_i^{(0)}P\right)~.
\end{align}
This construction explicitly shows that $a_{i}^{(0)}$ corresponds to the boundary Weyl connection, though it remains pure gauge in the bulk. 
It is also common in the literature to write the Weyl covariant derivative in terms of the Levi--Civita connection 
\begin{align}\label{Gzero}
    \Gamma_{(0)}^m{}_{ij}:= \frac12 h_{(0)}^{mn}\left(\partial_i h^{(0)}_{nj} + \partial_j h^{(0)}_{ni} - \partial_n h^{(0)}_{ij}\right)~.
\end{align}
The action of the Weyl covariant derivative on a scalar $S$ and vector field $V$ of Weyl weight $w$, can be obtained by expanding \eqref{Weylring},
\begin{align}
    \hat{\nabla}^{(0)}_iS={}&\nabla_i^{(0)}S+wa^{(0)}_iS\,,\\
    \hat{\nabla}^{(0)}_iV_j={}&\nabla_i^{(0)} V_j+(w+1)a^{(0)}_iV_j+a^{(0)}_jV_i-h^{(0)}_{ij}a^m_{(0)}V_m\,.
\end{align}
Using the aforementioned elements, one can construct Weyl--Riemann curvature $\hat{\cal R}^{i}_{(0)jmn}$, the Weyl--Ricci tensor $\hat{\cal R}_{ij}^{(0)}$, and scalar $\hat{\cal R}^{(0)}$~. For instance,
\begin{align}
    \hat{\cal R}^{(0)} = {\cal R}^{(0)} + 2\nabla^{(0)}_i a^i_{(0)}~,
\end{align}
where ${\cal R}^{(0)}$ is the scalar curvature of $h_{(0)}$ and $\nabla_i^{(0)}$ the boundary covariant derivative with respect to \eqref{Gzero}~. 

In the presence of a Weyl connection, the field theory gets a correction to the Weyl--Ward identity \cite{Ciambelli:2023ott, Jia:2023gmk} of the form\footnote{In \cite{Ciambelli:2019bzz} the Weyl covariant divergence of the current appears in the scale anomaly as in that case the Weyl weight of $J^i$ equals the spacetime dimension and one has $\nabla_i^{(0)}J^i = \hat\nabla_i^{(0)}J^i$~. }
\begin{align}\label{Weylanomaly}
    \frac{1}{\sqrt{-{\rm det}~\gamma^{(0)}}}\frac{\delta {\widehat{\cal A}}}{\delta \log\cB}=\langle T^{ij}\gamma^{(0)}_{ij} + \nabla_i^{(0)} J^i\rangle~,
\end{align}
where $\widehat{\cal A}$ is the celebrated Weyl anomaly \cite{Capper:1974ic,Capper:1975ig} and 
\begin{align}\label{Jigeneric}
    \langle J^i\rangle  = -\frac{1}{\sqrt{-{\rm det}~\gamma^{(0)}}}\frac{\delta I}{\delta a_i^{(0)}}~,
\end{align}
is the Weyl current operator. 
Then, the Weyl current contributes to the Weyl anomaly of the dual theory, adding a new term that makes the anomaly a Weyl covariant quantity, though the full stress tensor itself remains non-covariant. For more details on Weyl geometry see for instance \cite{Scholz:2017pfo,Wheeler:2018rjb,Ciambelli:2019bzz, Jia:2021hgy,Jia:2023gmk, Jia:2024ujz} and reference therein.

In the WFG gauge, the dual theory to pure gravity seems to have a trivial Weyl current \cite{Ciambelli:2019bzz}. Nonetheless, the holographic stress tensor now contains a new contribution coming from the coefficient $a_i^{(d-2)}$ appearing at order $z^{d-2}$ of the expansion of $a_i$. This term plays the role of the Weyl current, making the trace of the stress tensor Weyl covariant.
In the case of three-dimensional gravity, this coefficient corresponds exactly to the Weyl connection $a_i^{(0)}$ meaning that the Weyl current becomes non-trivial \cite{Ciambelli:2023ott}. We will demonstrate that this non-trivial Weyl current also emerges in the gFG gauge.

%%%%%%%%%%%%%%%%%%%%%%%%%%%%%%%%
%%%%%% Generalized FG %%%%%%%%%%
%%%%%%%%%%%%%%%%%%%%%%%%%%%%%%%%
\section{Generalized Gauge}\label{Sec:gFG}
%%%%%%%%%%%%%%%%%
%%%%%%%%%%%%%%%%%
%%%%%%%%%%%%%%%%%
Consider an asymptotically anti-de Sitter space written in the WFG gauge
\begin{align}\label{gFG}
    ds^2 ={}& \ell^2\left(\frac{dz}{z}-a_i dx^i\right)^2 +h_{ij}dx^idx^j \equiv \frac{\ell^2}{z^2}\left(dz^2 - 2z a_idx^idz\right) + \gamma_{ij}dx^i dx^j~,
\end{align}
where 
\begin{align}\gamma_{ij} := h_{ij} + \ell^2 a_i a_j~
\end{align}
is the induced metric at constant $z$, and assuming that $h_{ij}$ and $a_{i}$ admit a power series expansion 
\begin{align}\label{asybeh}
    h_{ij} = \frac{\ell^2}{z^2}\sum_{k\geq0} \left(\frac{z}{\ell}\right)^k h^{(k)}_{ij}~,\qquad a_i = \sum_{k\geq 0} \left(\frac{z}{\ell}\right)^{k-1}a_i^{(k-1)}~.
\end{align}
Note that the expansion for $a_{i}$ allows extra contributions compared to \cite{Ciambelli:2019bzz}. In this case, the presence of these new terms implies that, when solving Einstein's equations, the odd coefficients of the metric expansion are still present. Although the gauge remains to be WFG, the asymptotics are modified. For this reason, and to make a distinction with the falloff conditions considered in \cite{Ciambelli:2019bzz} and following \cite{Grumiller:2016pqb}, we will refer to it as the gFG gauge.

The boundary metric $\gamma_{ij}$ has the familiar FG expansion
\begin{align}
    \gamma_{ij} = \frac{\ell^2}{z^2}\left(\gamma^{(0)}_{ij} + \frac{z}{\ell}\gamma_{ij}^{(1)} + \frac{z^2}{\ell^2}\gamma_{ij}^{(2)}\right) + {\cal O}(z)~, \qquad \gamma^{ij} = \frac{z^2}{\ell^2}\left(\gamma^{ij}_{(0)} + \frac{z}{\ell}\tilde{\gamma}^{ij}_{(1)} + \frac{z^2}{\ell^2}\tilde{\gamma}^{ij}_{(2)} \right)+{\cal O}(z^5)
\end{align}
 but now the coefficients $\gamma_{ij}^{(k)}$ can be expressed in terms of $h_{ij}^{(k)}$ and $a_{i}^{(m)}$. For instance, the relevant coefficients read \footnote{We use the following convention for (anti-) symmetrization: $A_{(ij)}=\frac12\left(A_{ij}+A_{ji}\right)$ and $A_{[ij]}=\frac12\left(A_{ij}-A_{ji}\right)$}
\begin{align}
\gamma_{ij}^{(0)} ={}& h^{(0)}_{ij} + \ell^2 a_{i}^{(-1)}a^{(-1)}_{j}~, \\ 
\gamma_{ij}^{(1)} ={}& h^{(1)}_{ij} + 2\ell^2 a_{(i}^{(-1)}a^{(0)}_{j)}~, \\ 
\gamma_{ij}^{(2)} ={}& h^{(2)}_{ij} + 2\ell^2 a_{(i}^{(-1)}a^{(1)}_{j)} + \ell^2 a_i^{(0)}a_j^{(0)}~,  
\end{align}
where their corresponding inverses are
\begin{align}
    \gamma^{ij}_{(0)} ={}& h_{(0)}^{ij} - \ell^2 a^i_{(-1)}a^j_{(-1)}~, \\ \tilde{\gamma}^{ij}_{(1)} ={}& -\gamma_{(1)}^{ij}~, \\ \tilde{\gamma}^{ij}_{(2)} ={}& -\gamma^{ij}_{(2)} + \gamma^{im}_{(1)}\gamma_{(1)}^j{}_m~.
\end{align}
Under the Weyl transformation \eqref{WeylDiff}, the line element \eqref{gFG} transforms as
\begin{align}
    ds'^2 = \ell^2\left(\frac{dz'}{z} - a'_i(\cB z',x')dx'^i\right)^2 + h_{ij}(\cB z',x')dx'^idx'^j~,
\end{align}
where
\begin{align}\label{aprime}
    a_i' = a_i - \partial_i \log\cB~.
\end{align}
Analogously to the WFG gauge, one can check that all subleading terms in the expansion \eqref{asybeh} transform Weyl covariantly 
\begin{align}
    h_{ij}^{(k)}\to h'{}_{ij}^{(k)} = \cB^{k-2}h_{ij}^{(k)}~,\qquad a_i^{(k)} \to a'{}_i^{(k)} = \cB^{k} a_i^{(k)} - \delta_{k,0}\partial_i \log \cB~.
\end{align}
Once again, the leading term $a_i^{(0)}$ transforms inhomogeneously generating a boundary Weyl connection such as in the WFG case. 

Another interesting feature of the gFG gauge is that the boundary metric takes the form
\begin{align}
    \gamma^{(0)}_{ij}\equiv\lim_{z\to0}\frac{z^2}{\ell^2}\gamma_{ij} = h^{(0)}_{ij} +\ell^2 a_i^{(-1)}a_j^{(-1)}~.
\end{align}
In fact, this is a key distinction from the WFG gauge. In the WFG case, the boundary metric $\gamma_{ij}$ matches with $h_{ij}$ having no contributions from the asymptotic coefficients of the vector $a_i$~.
The inclusion of $a_{i}^{(-1)}$ not only modifies the boundary metric $\gamma_{ij}^{(0)}$, but also the solutions to Einstein's equations allowing odd coefficients of the induced metric $h_{ij}^{k}$ to be non-vanishing.  
Later, we will see that solving Einstein equations allows all subleading coefficients $h_{ij}^{(k)}$ to be expressed in terms of $h_{ij}^{(0)}$ and its derivatives. In contrast, all coefficients of the vector expansion $a_i^{(k)}$ remain arbitrary.
Thus, this highlights a subtle difference between the FG and the gFG gauge: in the gFG gauge, the dynamics alone do not fully determine the geometry solely through the boundary metric. Instead, to fully reconstruct the bulk structure in terms of the induced metric $\gamma_{ij}$, it is also necessary to specify the value of all coefficients of the asymptotic expansion of the vector $a_{i}$. 

\paragraph{Field equations.}
We now follow to solve Einstein's equations
\begin{align}
    {\cal E}_{\mu\nu} := R_{\mu\nu} -\frac12 R g_{\mu\nu} + \Lambda g_{\mu\nu} = 0~
\end{align}
order by order in the holographic coordinate $z$ to obtain the values of the expansion coefficients in terms of the undetermined boundary metric $\gamma_{(0)}$ and vector coefficients $a_i^{(k)}$~. At leading order $\mathcal{O}(\nicefrac{1}{z^2})$~, the ${\cal E}_{zz}$ components gives the following normalization for the Weyl vector $a_i^{(-1)}$ as the one of a null vector respect to $\gamma_{(0)}$, namely
\begin{align}
    a_i^{(-1)}a^{(-1)}_j \gamma^{ij}_{(0)} = 0~.
\end{align}
The next-to-leading order of the boundary components of the field equations ${\cal E}_{ij}$ gives
\begin{align}
    h_{ij}^{(1)} = -2\ell^2\hat\nabla_{(i}a^{(-1)}_{j)}~,
\end{align}
which, as we previously mentioned, does not appear in the FG gauge nor the WFG gauge for pure Einstein gravity.  This shows a deviation of the gFG gauge with respect to the WFG gauge as now there are odd terms in the power series expansion appearing due to the new subleading vector coefficient $a^{(-1)}_i$.
Finally, the zeroth order in the field equations gives
\begin{align}
    {\rm Tr}~h_{(2)} = -\frac{\ell^2}{2}\hat{\cal R}^{(0)}-\ell^2 h^{(2)}_{ij}a^i_{(-1)}a^j_{(-1)}~,
\end{align}
where the trace is taken with the boundary metric, i.e., ${\rm Tr}~h_{(2)} = \gamma^{(0)}_{ij}h_{(2)}^{ij}$~.
It is worth noticing that when the subleading vector coefficient $a^{(-1)}_i$ is turned off, all the expressions above reduce to known results of the WFG gauge \cite{Ciambelli:2019bzz, Jia:2021hgy, Jia:2023gmk}. Similarly, the FG gauge \cite{deHaro:2000vlm} is also recovered by setting $a_{i}$ to zero. This shows that the gFG gauge contains both the FG and WFG gauges as special limiting cases.

\paragraph{Renormalized action.} 
Three-dimensional Einstein gravity in an AlAdS background is described in terms of the Einstein--Hilbert action with negative cosmological constant and the Gibbons--Hawking--York boundary term 
\begin{align} \label{AdS3action}
    I_{\rm EH} = \frac{1}{16\pi G}\int_{\cal M}d^3x \sqrt{-g}\left(R+\frac{2}{\ell^{2}}\right) + \frac{1}{8\pi G}\int_{\partial \cal M}d^2 x\sqrt{-\gamma}K~, 
\end{align}
where $K = \gamma^{ij}K_{ij}$ is the trace of the extrinsic curvature. Let us write the bulk metric in an ADM-like frame
\begin{align}
    ds^2 = N^2 dz^2 + \gamma_{ij}\left(dx^i + N^i dz\right)\left(dx^j + N^j dz\right)~,
\end{align}
where we have chosen a decomposition in the $z-$direction, $N_{i}$ is the shift function, and $N$ is the lapse function. In these coordinates, a normal vector to the boundary is given as
\begin{align}\label{normal}
    n = \frac{1}{N}\left( N^i \partial_i - \partial_z\right)~.
\end{align}
The extrinsic curvature is obtained as the Lie derivative of the metric along the normal vector $n$,
\begin{align}
    K_{ij} = \frac{1}{2}{\cal L}_n g_{ij} = -\frac{1}{2N}\left( \partial_z \gamma_{ij} - \widebar\nabla_i N_j - \widebar\nabla_j N_j\right)~.
\end{align}
Comparing with \eqref{gFG}, we find that the shift and the lapse function can be expressed as
\begin{align}
    N_i ={}& -\frac{\ell^2}{z}a_i~,
    \\ N^2 ={}& \frac{\ell^2}{z^2}- \gamma^{ij}N_i N_j = \frac{\ell^2}{z^2}\left(1-\ell^2 a_i a^i\right)~.
\end{align}

Due to the infinite volume of AdS spacetimes, the action encounters infrared (IR) divergences. A systematic approach to address this issue goes as follows: First, one evaluates the action near the boundary using the Fefferman-Graham expansion, introducing an IR cutoff at $z=\delta <<1$. This separates finite and divergent contributions. Based on the explicit form of the divergent terms, covariant counterterms, that depend solely on intrinsic quantities of the boundary, are constructed. By adding these covariant boundary counterterms to the action, the IR divergences are cancelled, allowing the limit $\delta\to0$ to be taken at the end of the calculation. This procedure, known as \textit{holographic renormalization} \cite{deHaro:2000vlm, Skenderis:2002wp, Henningson:1998gx}, allows for the extraction of physical observables and correlation functions in the field theory. For three dimensions, the boundary counterterms were obtained in the FG gauge in \cite{Balasubramanian:1999re}, where the action takes the form
\begin{align}
    I_{\rm FG} = I_{\rm EH} + I_{\rm ct}~,
\end{align}
with
\begin{align}
    I_{\rm ct} = - \frac{1}{8\pi  G\ell}\int_{\partial\cal M} \sqrt{-\gamma}~.
\end{align}
%%%
In the WFG gauge, an additional counterterm of the form
\begin{align}\label{k2ciambelli}
    I_{a^2} = \frac{1}{16\pi G}\int_{\partial\cal M}d^2x \sqrt{-\gamma} \gamma^{ij}a_i a_j~
\end{align}
was introduced in \cite{Ciambelli:2023ott} to ensure the finiteness of the on-shell action. In fact, this boundary term is finite on-shell and therefore, gives a finite contribution to the total energy of the system. Moreover, the counterterm modifies the stress tensor of the boundary field theory adding a term proportional to $a_{(0)}^2$.  In addition, there is a holographic Weyl current associated with the variations of the action with respect to $a_{i}^{(0)}$. The Weyl current is proportional to the connection $a_{(0)}^i$, so it is not a Weyl tensor. 
Despite this, in \cite{Ciambelli:2023ott}, the Weyl current is treated as a Weyl vector by considering its tensorial weight as its Weyl weight. Using this, they construct a Weyl covariant derivative of the current. Then, in the holographic stress tensor, there is a new contribution coming from \eqref{k2ciambelli} that gives the extra terms to construct such a Weyl covariant derivative of the Weyl current. 
As the tensorial weight of the current equals the boundary dimension, the covariant divergence of this object does not get a modification due to the extra Weyl connection.

We proceed to study the holographic renormalization scheme within the gFG gauge. As we will show later, the same Weyl current found for the WFG gauge in \cite{Ciambelli:2023ott} emerges in the gFG by varying the effective boundary action \eqref{QEA} without requiring any additional finite contributions. Moreover, we will not treat this current as a Weyl vector, since it is proportional to the boundary Weyl connection.

In the gFG gauge, the crossed terms in the metric contribute to the second-order pole that defines the boundary conformal structure. These new divergences -- subleading relative to the bulk divergences -- are induced directly by the GHY term. The counterterm \eqref{k2ciambelli} is unsuitable in this context, as it leads to ${\cal{O}}(\nicefrac{1}{z})$ divergences in the on-shell action. Consequently, new covariant counterterms are necessary. To identify these, we analyze the asymptotic behaviour of the action \eqref{AdS3action}, thereby allowing us to construct appropriate covariant counterterms.
For the bulk metric determinant, the asymptotic expansion is given by
\begin{align}
    \sqrt{-g} = \sqrt{-\gamma_{(0)}}\left(\frac{\ell^3}{z^3} + \frac{\ell^2}{2z^2}X^{(1)} + \frac{\ell}{2z}X^{(2)} + {\cal O}(1)\right)~,
\end{align}
where $X^{(1)}$ and $X^{(2)}$ are functions to be determined in terms of the boundary data. Substituting the expansion into the on-shell action, we find
\begin{align}
    X^{(1)} = -2\ell^2\hat\nabla_i^{(0)} a^i_{(-1)}~, \qquad X^{(2)} = -\frac{\ell^2}{2}\hat{\cal R}~.
\end{align}
The expansion for the square root of the boundary metric determinant is
\begin{align}
    \sqrt{-\gamma} = \sqrt{-\gamma_{(0)}}\left(\frac{\ell^2}{z^2} + \frac{\ell}{z}W^{(1)} + W^{(2)} + {\cal O}(z)\right)~,
\end{align}
with 
\begin{align}
    W^{(1)} = \frac12 X^{(1)}+\ell^2 a_i^{(0)}a^i_{(-1)}~
\end{align}
and 
\begin{align}
    W^{(2)}=\frac12 X^{(2)}-\frac{\ell^2}{2}h^{(2)}_{ij}a^{i}_{(-1)}a^{j}_{(-1)}+\frac{\ell^2}{2}a^{(0)}_ia^{i}_{(0)}+\ell^2a^{(-1)}_ia^{i}_{(1)}~.
\end{align}

The asymptotic behaviour of the trace of the extrinsic curvature reads
\begin{align}
    K = K^{(0)} + zK^{(1)} +z^2 K^{(2)} + {\cal O}(z^3)~,
\end{align}
with 
\begin{align}
    K^{(0)} = \frac{2}{\ell}~, \qquad
    K^{(1)} = 0~,\qquad K^{(2)} = -\frac{1}{\ell^3}X^{(2)} - \frac{1}{\ell}\nabla_i^{{(0)}} a^i_{(0)}~,
\end{align}

where $\nabla_i^{(0)}$ is the Levi--Civita covariant derivative with respect to $\gamma_{(0)}$~. 
Next, consider the renormalized action now including an extra boundary term $\Upsilon$, so that it reads
\begin{align}
    I_{\rm ren} = \frac{1}{16\pi G}\int_{\cal M}d^3x \sqrt{-g}\left(R+\frac{2}{\ell^2}\right) + \frac{1}{8\pi G}\int_{\partial\cal M}d^2x \sqrt{-\gamma}\left( K -\frac{1}{\ell} + \Upsilon\right)~.
\end{align}
We assume that $\Upsilon$ has an asymptotic expansion of the form
\begin{align}
    \Upsilon = z \Upsilon^{(1)} + z^2 \Upsilon^{(2)} + {\cal O}(z^3)~,
\end{align}
where $\Upsilon^{(1)}$ is determined by analyzing the new divergences appearing in the GHY term. 
Combining all the terms and integrating the bulk, the expansion of the total action becomes
\begin{align}
    I_{\rm ren} = -\frac{1}{8\pi G}\int_{\partial\cal M} d^2x\sqrt{-\gamma_{(0)}}\left[\frac{1}{z}\left(X^{(1)} - W^{(1)} - \ell^2 K^{(1)} - \ell^2 \Upsilon^{(1)}\right) + X^{(2)}\log z + {\cal O}(1)\right]~.
\end{align}
The logarithmic divergence accounts for the term responsible for the holographic Weyl anomaly \cite{Skenderis:2002wp}, and it must be cancelled by an additional counterterm that is proportional to the Ricci scalar in the FG gauge \cite{deHaro:2000vlm}. Thus, it does not contribute to the holographic stress tensor in $d=2$ as it is a topological invariant\footnote{Later in this section, we see that in the gFG gauge, the $X^{(2)}$ factor becomes proportional to the boundary Weyl--Ricci scalar. However, the difference between the Ricci scalar and the Weyl--Ricci scalar is merely a total derivative of the Weyl connection. As a result, this counterterm does not alter the holographic stress tensor in the gFG gauge either.}. 

To cancel the $\nicefrac{1}{z}$ divergence, the leading order of  the counterterm $\Upsilon$ must be
\begin{align}\label{beta1}
    \Upsilon^{(1)} ={}& \frac{1}{\ell^2}X^{(1)} - \frac{1}{\ell^2}W^{(1)} - K^{(1)} \nonumber \\ ={}& -\nabla_i^{(0)} a^i_{(-1)}~. 
\end{align}

A natural covariant expression (with respect to $\gamma$) that recovers \eqref{beta1} at first order is
\begin{align}\label{tildebeta}
    \widetilde{\Upsilon} = -\widebar\nabla_i a^i~.
\end{align}
Although this term is a total derivative, it could potentially become relevant in the presence of spacetimes with topological singularities. In that case, this term corresponds to a codimension-two counterterm, recently referred to as a \textit{corner term} in \cite{Ciambelli:2023ott}, which is also important for extracting the gravitational symplectic potential\footnote{See \cite{Ciambelli:2022vot} for a review on corner terms and its implication on conserved charges and asymptotic symmetries.}. In the WFG gauge, this corner term is finite, covariant with respect to boundary diffeomorphisms, and cancels logarithmic divergences. However, in the present context, the use of such a counterterm is needed to cancel new subleading divergences but introduces new finite contributions to the on-shell action when it is not trivial as one could check by evaluating \eqref{tildebeta} in the specific example of the accelerating BTZ black hole (see \autoref{Sec:Examples}). As a result, the holographic energy is shifted, and gives the wrong thermodynamic interpretation to the Euclidean on-shell action, as we show later in \autoref{Sec:Examples}. We encounter a similar issue in \autoref{App:pneq1} when considering the most general set of AdS$_3$ boundary conditions, where, moreover, no covariant analogue exists. 

%%%%%%
We propose the following counterterm instead
\begin{align}\label{cod2counter}
    \Upsilon = \gamma^{ij}\widebar\nabla_i n_j~,
\end{align}
where $n_i$ is the boundary component of the normal defined in \eqref{normal}, i.e.,
\begin{align}
    n^i = \frac{N^i}{N} \equiv -\frac{\ell a^i}{\sqrt{1 - \ell^2 a_i a^i}}~.
\end{align}
Similar to \eqref{tildebeta}, this counterterm is a total derivative. However, unlike \eqref{tildebeta}, it does not affect the total energy of the system but instead addresses the subleading divergences required to regulate those arising from the GHY term.  Furthermore, the applicability of \eqref{cod2counter} is broader than that of \eqref{tildebeta}. In \autoref{App:pneq1}, we employ this counterterm for the most general AlAdS spacetime, extending beyond the gFG gauge. It becomes especially relevant in topologically non-trivial configurations, such as the accelerating BTZ black hole in 2+1 dimensions, as demonstrated in \autoref{Sec:Examples}. 

Using this counterterm, the total renormalized action in the gFG gauge is
\begin{align}\label{Sren}
I_{\rm ren} = \frac{1}{16\pi G}\int_{\cal M}d^3x \sqrt{-g}\left(R+\frac{2}{\ell^{2}}\right) + \frac{1}{8\pi G}\int_{\partial\cal M}d^2x \sqrt{-\gamma}\left(K - \frac{1}{\ell} + \widebar\nabla_i n^i\right)~,  \end{align}
and the holographic quantum effective action that generates connected diagrams of the dual field theory is
\begin{align}\label{QEA}
    I_{\rm ren} = \frac{1}{8\pi G}\int_{\partial \cal M}d^2x\sqrt{-\gamma_{(0)}}{}&\left[ -\frac{1}{2\ell}X^{(2)}-\frac{\ell}{2}h^{(2)}_{ij}a^{i}_{(-1)}a^{j}_{(-1)} +\ell a_{i}^{(1)}a_{(-1)}^i + \frac{\ell}{2}a_i^{(0)}a^i_{(0)}+\right.\nonumber
    \\
    {}&\left.- \nabla_i^{(0)}\left(2\ell a_{(0)}^i + \ell^3a_j^{(0)}a_{(-1)}^j a^i_{(-1)}\right)\right]~.
\end{align}

Now, we find the boundary Weyl current by considering variations of the renormalized action with respect to the Weyl connection $a^{(0)}_{i}$ (see \eqref{Jigeneric}), viz. 
\begin{align}\label{JWeyl}
    J^i = -\frac{1}{\sqrt{-\gamma_{(0)}}}\frac{\delta I_{\rm ren}}{\delta a^{(0)}_i} = -\frac{\ell}{8\pi G}a^i_{(0)}~,
\end{align} 
matching the result of \cite{Ciambelli:2023ott} for the WFG gauge. As previously mentioned, this current is not a Weyl tensor, as it is directly proportional to the Weyl connection. Therefore, applying a Weyl covariant derivative to it is not well-defined. Next, we will show that the trace of the stress tensor can be decomposed into a Weyl covariant component and an inhomogeneous term, the latter given by the boundary covariant divergence of the Weyl current.

%%%%%%%%%%%%%%%%%%%%%%%%%%%%%%%%%%%%%%
%%%%%%%%%%%%%%%%%%%%%%%%%%%%%%%%%%%%%%
%%%%%%%%%%%%%%%%%%%%%%%%%%%%%%%%%%%%%%
%%%%%%%%%%%%%%%%%%%%%%%%%%%%%%%%%%%%%%
%%%%%%%%%%%%%%%%%%%%%%%%%%%%%%%%%%%%%%

\paragraph{Holographic Weyl anomaly.} 

Since the action is modified only by a total derivative, the holographic stress tensor in the gFG frame can be obtained from the $z \to 0$ limit of the quasilocal stress tensor
\begin{align}\label{quasi_local}
    \tau_{ij} = \frac{1}{8\pi G}\left(K_{ij}-\gamma_{ij}K + \frac{1}{\ell}\gamma_{ij}\right)~,
\end{align}
as done in \cite{Balasubramanian:1999re}. Consequently, the holographic stress tensor \cite{deHaro:2000vlm} is 
\\
\begin{align}
\langle T_{ij}\rangle:={}& \frac{2}{\sqrt{-\gamma_{(0)}}}\frac{\delta I_{\rm ren}}{\delta \gamma^{ij}_{(0)}} = \lim_{z\to0}\tau_{ij}={}\frac{1}{8\pi G}\left(-\frac{1}{\ell}\gamma^{(2)}_{ij}-\ell^2\gamma^{(0)}_{ij}+K^{(2)}_{ij}\right)
   \nonumber\\
    ={}&\widehat{T}_{ij} - \frac{\ell}{8\pi G}\left[\nabla^{(0)}_{(i}a_{j)}^{(0)} - \gamma_{ij}^{(0)}\nabla^{(0)}_l a^l_{(0)}+a^{(0)}_ia^{(0)}_j-\frac12\gamma_{ij}a^{(0)}_la^{l}_{(0)}\right]\nonumber
    \\
    {}&+\ell^2a^{l}_{(-1)}\left(a^{(-1)}_{(i}\nabla^{(0)}_{l}a^{(0)}_{j)}-a^{(-1)}_{(i}\nabla^{(0)}_{j)}a^{(0)}_l\right)\,,
\end{align}
where 
\begin{align}
    \widehat{T}_{ij}={}&-\frac{1}{8\pi G}\left[\frac{1}{\ell}\left(h^{(2)}_{ij}+\frac{\ell^2}{2}\gamma^{(0)}_{ij}h^{(2)}_{lk}a^{l}_{(-1)}a^{k}_{(-1)}-\gamma^{(0)}_{ij}X^{(2)}\right)+2\ell^2\left(a^{(-1)}_{(i}a^{(1)}_{j)}-\frac12\gamma^{(0)}_{ij}a^{l}_{(-1)}a^{(1)}_{l}\right)\right.\nonumber
    \\
    {}&-\left.\ell^2a^{l}_{(-1)}\left(\hat{\nabla}_{(i}h^{(1)}_{j)l}-\frac12\hat{\nabla}_l h^{(1)}_{ij}\right)\right]
\end{align}
is the Weyl covariant part of the holographic stress tensor.

Using \eqref{JWeyl} we can rewrite the stress tensor as
\begin{align}\label{holo_stress2}
    \langle T_{ij}\rangle={}&\widehat{T}_{ij}+\nabla^{(0)}_{(i}J_{j)}-\gamma^{(0)}_{ij}\nabla^{(0)}_lJ^{l}-\frac{8\pi G}{\ell}\left(J_{i}J_{j}-\frac12\gamma^{(0)}_{ij}J^{l}J_{l}\right)\nonumber
    \\
    {}& +\ell^2a^{l}_{(-1)}\left(a^{(-1)}_{(i}\nabla^{(0)}_{l}J_{j)}-a^{(-1)}_{(i}\nabla^{(0)}_{j)}J_l\right)\,.
\end{align}

In this way, the holographic Weyl anomaly is given by 
\begin{align}\label{tildeTij}
    \widehat{\mathcal A} = \langle \widetilde{T}^{ij}\gamma^{(0)}_{ij}\rangle \equiv \langle T^{ij}\gamma^{(0)}_{ij} + \nabla_i^{(0)} J^i\rangle~,
\end{align}
where the trace of the holographic stress tensor is
\begin{align}\label{Trace_T}
    \langle T^{i}{}_i\rangle= \widehat{T}^{i}{}_i-\nabla^{(0)}_i J^i
     = \frac{1}{8\pi G\ell}X^{(2)}-\nabla^{(0)}_iJ^{i}\,.
\end{align}
Therefore, the trace of \eqref{tildeTij} gives a Weyl covariant holographic trace anomaly
\begin{align}\label{tildeTii}
    \widehat{\cal A} = \widehat{T}^i{}_i \equiv -\frac{\ell}{16\pi G}\hat{\cal R}^{(0)}~,
\end{align}
obtaining a Weyl covariant conformal anomaly with the correct Brown--Henneaux central charge \cite{Brown:1986nw}, $c=\nicefrac{3\ell}{2G}$~, as in \cite{Jia:2021hgy,Ciambelli:2023ott}. 

%%%%%%%%%%%%%%%%%%%%%%%%%%%%%%%%
%%%%%%%%% Examples %%%%%%%%%%%%%
%%%%%%%%%%%%%%%%%%%%%%%%%%%%%%%%
\section{Examples}\label{Sec:Examples}
%%%%%%%%%%%%%%%%%
%%%%%%%%%%%%%%%%%
%%%%%%%%%%%%%%%%%
\paragraph{Integrable models.} In \cite{Cardenas:2021vwo} it is shown that the asymptotic dynamics of three-dimensional Einstein gravity with negative cosmological constant can describe an Ablowitz--Kaup--Newell--Segur (AKNS) integrable hierarchy, which characterizes a wide class of integrable systems in two dimensions \cite{Ablowitz:1973zz}. To achieve that, the Brown--Henneaux boundary conditions \cite{Brown:1986nw} are relaxed to include more general bulk geometries. Particularly, it shows the necessity of cross terms between the holographic and boundary coordinates, generating a boundary Weyl connection and higher terms in the expansion, i.e., $a_i^{(n)}$. for $n\geq0$. This shows how the WFG gauge can play an important role in the understanding of two-dimensional holographic theories. Nonetheless, there is no $a_i^{(-1)}$ term in their examples. Along these lines, in \cite{Cardenas:2024hah}, it has been proposed that studying the two-dimensional version of the gFG gauge \cite{Grumiller:2013swa}, the gauge formulation of Jackiw--Teitelboim gravity contains an infinite set of commuting Hamiltonians associated with the Korteweg--de Vries (KdV) hierarchy. Just as in the three-dimensional case, the bulk solution now contains a $a_i$ vector that only contains $a_i^{(0)}$ in its expansion such that the dual theory contains a Weyl connection. Both examples indicate that is possible to get holographic CFTs that are both integrable and Weyl covariant within the scheme studied in this article.

\paragraph{Accelerating black holes.} Although three-dimensional gravity lacks dynamical degrees of freedom, it is well-known that the theory admits a non-trivial black hole solution \cite{Banados:1992gq} in the presence of a negative cosmological constant. It is also possible to construct non-trivial solutions that include topological defects, such as domain walls. This is the case for accelerating configurations \cite{Anber:2008zz, Astorino:2011mw, Xu:2011vp, Arenas-Henriquez:2022www}, whose thermodynamics and holographic properties have been recently  investigated in \cite{Arenas-Henriquez:2023vqp, Lei:2023mqx, Xu:2023tfl, Kim:2023ncn, Tian:2023ine,Tian:2024mew,Fontana:2024odl,Javed:2024nqm, Kim:2024ncx, Kim:2024dbj, Bunney:2024xic} . More precisely, accelerating black holes correspond to a one-parameter extension of the BTZ black hole. The entire family of solutions is described by the line element
\begin{align} \label{accfamily}
    ds^2 = \frac{1}{\Omega(r,\phi)^2}\left(-f(r) dt^2 + \frac{dr^2}{f(r)} + r^2 d\phi^2\right)~,
\end{align}
where both, the conformal factor $\Omega(r,\phi)$, which depends on the radial $r$ and angular $\phi$ coordinates, and $f(r)$, the metric function, are specified by the particular accelerated solution under consideration \cite{Arenas-Henriquez:2022www}. Just as in the four-dimensional counterpart (see for instance \cite{Dias:2002mi}), the black hole is accelerated due to a domain wall (codimension-one topological defect) that generates a conical deficit. In particular, the accelerating BTZ black hole is given by
\begin{align}
    f(r) = \frac{r^2}{\ell^2} - m^2\left(1-{\cal A}^2r^2\right)~,\qquad \Omega = 1+{\cal A}r \cosh(m\phi)~,
\end{align}
where $\phi\in (-\pi,\pi)$~, and $m$ is a parameter that defines the location of the domain wall. The domain wall acts as an internal boundary that symmetrically separates the spacetime into two regions. This induced boundary has a non-trivial jump in the extrinsic curvature; hence, the Israel junction condition determines the energy density of the domain wall, which corresponds to its tension
\begin{align}\label{tension}
    \mu = - \frac{m\cA \sinh(m\pi)}{4\pi G}~.
\end{align}

The particular form of the metric \eqref{accfamily} complicates the identification of the holographic coordinate, as the conformal boundary corresponds to the surface defined by $\Omega(r,\phi) = 0$ rather than a single point. For the accelerating BTZ black hole, we have that the conformal boundary corresponds to the surface parametrized by $r = ({\cal A}\cosh(m\phi))^{-1}$ rather than the usual $r = \infty$ of the BTZ black hole. This feature is also shared with its four-dimensional counterpart and in the presence of additional matter fields \cite{EslamPanah:2022ihg, Barrientos:2022bzm, Cisterna:2023qhh, Kubiznak:2024ijq, Kim:2024ncx, Barrientos:2024umq}.
Consequently, posing \eqref{accfamily}, or more generally, any accelerating solution in the FG gauge requires a more involved procedure, in which both radial and angular coordinates are expanded in terms of two new coordinates and arbitrary functions of these coordinates. These functions are determined by requiring the gauge conditions of the FG frame, specifically requiring the metric to be devoid of cross terms between the holographic and boundary coordinates and $g_{zz} = \nicefrac{1}{z^{2}}$; see for instance \cite{Anabalon:2018qfv,Anabalon:2018ydc} for a more detailed discussion in four-dimensions. After achieving the FG gauge, the boundary metric is not fully determined as this procedure leaves the $h_{(0)}$ coefficient of the asymptotic expansion unfixed, explicitly illustrating the conformal freedom of the boundary. In three dimensions, the two-dimensional boundary CFT suffers from a Weyl anomaly, which is independent of the choice of conformal representative of the boundary metric, while the stress tensor is a quasi-primary restricting the computation of observables unless the representative is consistently fixed. This situation poses a significant issue, as there appears to be a lack of consensus regarding a procedure for concretely choosing a conformal representative. For further discussions, see \cite{Papadimitriou:2005ii,Arenas-Henriquez:2022www, Arenas-Henriquez:2023hur, Cisterna:2023qhh, Kim:2023ncn, Liu:2024fvq}. Therefore, many of the holographic and thermodynamic properties of accelerating black holes still remain unclear in the community.\\
An alternative approach was employed in \cite{Hubeny:2009kz, Cassani:2021dwa, Arenas-Henriquez:2023hur}, where instead, one can consider a simple coordinate transformation that renders the conformal factor $\Omega$ to be a function of a single new coordinate, say $z$. Such transformation introduces a cross term between the new holographic coordinate and a boundary coordinate in the line element, which means that it does not correspond to an FG coordinate system beyond the leading order. In this case, the boundary metric is fully determined and therefore, it is a more suitable approach for conducting a holographic analysis. Here, we will show that in fact, a similar coordinate transformation leads to the gFG gauge with a non-trivial Weyl connection. 

Following \cite{Hubeny:2009kz,Cassani:2021dwa,Arenas-Henriquez:2023hur}, let us consider for a moment the coordinate transformation
\begin{align}
    \frac{1}{r} = z - {\cal A}\cosh(m\phi)~
\end{align}
for the accelerating BTZ black hole. The new line element now takes the form of \eqref{gFGpneq1}. Although such transformation induces a cross term that induces an $a_{(-1)}$ coefficient, the boundary Weyl connection is trivial. A more refined coordinate transformation that induces a full expansion for the $a_i$ vector is \footnote{In general, one could consider a transformation of the form
\begin{align}
    \frac{1}{r} = \sum_{m\geq0}^\infty \zeta_{(m)}(x^i)z^m~,
\end{align}
fixing only $\zeta_{(0)} = -{\cal A}\cosh(m\phi)$ and $\zeta_{(1)} = 1~$. The metric then would satisfy all the requirements of a gFG system. Nonetheless, as we show later, only the quadratic coefficient is necessary to generate a boundary Weyl manifold for the accelerating BTZ black hole.}
\begin{align}\label{magictransf}
    \frac{1}{r} = z - {\cal A}\cosh(m\phi) + \zeta \ell z^2~.
\end{align}
Thus, the spacetime is now in the gFG coordinate system: it is an asymptotically locally AdS space with conformal boundary at $z=0$~, and it features a non-trivial boundary Weyl connection with an infinite asymptotic expansion \eqref{asybeh}. The induced vector becomes
\begin{align}\label{aiaccelerating}
   a_i = \frac{\mathcal{A} m \ell  \sinh (m \phi)}{z+2 \zeta   \ell z^2}\delta^\phi_i~,
\end{align}
and the leading terms in the asymptotic expansion are
\begin{align}
    a^{(-1)}_\phi = m{\cal A}\sinh(m \phi)~,\qquad a^{(0)}_\phi = 2m{\cal A}\ell\zeta \sinh(m\phi)~, \qquad a^{(1)}_\phi = 4m{\cal A}\ell^2\zeta^2\sinh(m\phi)~. 
\end{align}
Although in principle arbitrary, the $\zeta$ coefficient allows having a non-trivial expansion that incorporates a holographic Weyl connection $a_{(0)}$~. In fact, a more general transformation would allow $\zeta = \zeta(t,\phi)$ which induces an extra time component in the vector $a_{i}$ with its expansion starting at $\mathcal{O}(z)$. The resulting coefficients in expansion would result in more complicated combinations involving derivatives of $\zeta$ with both, angular and time components. For simplicity, we restrict ourselves to the case $\zeta= \rm{const}$, which is the minimal setting where the vector $a_{i}$ has a full asymptotic expansion, generating a boundary Weyl geometry. 

%\FD{write better} One could consider a more general transformation with $\zeta = \zeta(t,\phi)$ to induce an expansion for the time component of the vector starting at linear order in $z$, and the coefficients in the angular components would depend also be more complicated and depend on derivatives of the coefficients, i.e. $\partial_\phi \zeta$~. For the sake of simplicity, we focus on the simplest setup that gives us a full expansion of the vector, generating a boundary Weyl geometry. 

The boundary metric
\begin{align}
    ds_{(0)}^2 = \gamma^{(0)}_{ij}dx^idx^j = -\left(1-\mathcal{A}^2 m^2 \ell ^2 \sinh ^2(m \phi)\right)\frac{dt^2}{\ell^2} + \frac{d\phi^2}{1-\mathcal{A}^2 m^2 \ell ^2 \sinh ^2(m \phi)}
\end{align}
has a non-constant but always positive scalar curvature
\begin{align}
    {\cal R}^{(0)} = 2 \mathcal{A}^2 m^4 \ell ^2 \cosh (2 m \phi)~.
\end{align}
Note that the boundary metric, and thus its scalar curvature, does not depend on $\zeta$. The associated holographic stress tensor \eqref{holo_stress2} is also independent of $\zeta$, matching the one obtained in \cite{Arenas-Henriquez:2023hur}. Hence, while the $\zeta$ coefficient remains arbitrary, it does not influence the boundary observables. However, it does play a role in the holographic Weyl anomaly as  \eqref{tildeTii} includes contributions from the Weyl current. In this case, the Weyl current is given by
\begin{align}
    J^i =   \frac{\mathcal{A} m \ell^4}{2\pi G} \left(1-\mathcal{A}^2 m^2 \ell ^2 \sinh ^2(m \phi)\right) \zeta \sinh (m \phi) \delta_\phi^i~,
\end{align}
which ensures that the trace of the stress tensor is invariant under Weyl transformations (see \eqref{Trace_T}) and recovers the Brown--Henneaux central charge.

As previously mentioned, the accelerating BTZ black hole corresponds to a one-parameter extension of the BTZ black hole that possesses a domain wall in its interior extending from the black hole horizon to the conformal boundary. Geometrically, the spacetime splits into two $\mathbb{Z}_2$-symmetric regions. This region can be interpreted as an internal boundary for which the action should be supplemented with an extra GHY term that will ensure a well-posed variational principle producing the Israel equations. Then, the intersection between the domain wall and the conformal boundary creates a singular point that is in fact, regulated by the GHY term of the internal boundaries and energy-momentum of the domain wall. As commented in \cite{Tian:2023ine, Tian:2024mew}, this intersection point generates a boundary in boundary CFT allowing the use of AdS/BCFT correspondence \cite{Takayanagi:2011zk, Fujita:2011fp} techniques to study three-dimensional accelerating black holes.

In the present case, we notice that a direct evaluation of the on-shell renormalized action \eqref{Sren} has a non-trivial contribution from the codimension-two term \eqref{cod2counter}. This counterterm \eqref{cod2counter} 
has support in the intersection point between the domain wall and the conformal boundary, producing the necessary contribution that cancels the boundary divergence\footnote{This also happens in similar scenarios, for example, a BTZ black hole with an end-of-the-world brane intersecting the asymptotic boundary. Recently, it has been shown that the Euclidean action of such spacetimes, computed in the FG frame, leads to inconsistent results, see \cite{Tian:2024fmo,Lai:2024inw} for further details.}.  One can check that this singularity disappears in the zero-acceleration limit ${\cal A}\to0$ which removes the domain wall; in other words, its tension \eqref{tension} totally vanishes. 

Then, we find that the Euclidean version of the on-shell action \eqref{Sren} is finite and satisfies the quantum statistical relation, viz. 
\begin{align}
    I_{\rm reg} = \beta M - S~,
\end{align}
where 
\begin{align}
    M = \int d\phi \sqrt{\gamma_{(0)}}\langle T^t{}_t\rangle = \frac{m^2\left[2\pi(2+m^2{\cal A}^2\ell^2) - 3m{\cal A}^2\ell^2\sinh(2\pi m)\right]}{32\pi G} 
\end{align}
is the holographic mass obtained in \cite{Arenas-Henriquez:2023hur}, 
\begin{align}
    \beta = \frac{m\sqrt{1+m^2{\cal A}^2\ell^2}}{2\pi \ell}
\end{align}
is the inverse of the horizon temperature, and
\begin{align}
    S = \frac{\ell}{G}\arctanh\left[\left(\sqrt{1+m^2{\cal A}^2\ell^2}-m{\cal A}\ell\right)\tanh\left(\frac{m\pi}{2}\right)\right]
\end{align}
is the Bekenstein--Hawking entropy of the black hole horizon. 

As we have seen, the gFG gauge gives a suitable laboratory to study the thermodynamics of accelerating black holes and holographic properties. Particularly, the coordinate transformation \eqref{magictransf} is suitable to construct a Weyl covariant boundary theory. 
The procedure does not depend on the number of dimensions, and therefore it can be used to study four-dimensional accelerating black holes which have been proposed to be holographically dual to strongly and weakly coupled field theory plasma living on a black hole background \cite{Hubeny:2009kz}.

\section{Discussion and Outlook} \label{Sec:Discussions}
\paragraph{Summary.} In this work, we have studied the generalized Fefferman-Graham gauge in the metric formalism and its connection with boundary Weyl structures in three-dimensional gravity. We began by introducing the concept of the Weyl manifold arising at the boundary of AdS space in the WFG gauge and some of their essential properties, setting up the geometric framework. Following this, we defined the gFG gauge which is nothing else than the WFG gauge with modified falloff given by the asymptotic expansion for the boundary metric $\gamma_{ij}$ which is decomposed into $h_{ij}$ and the vector $a_{i}$. The leading component of the $a_{i}$ vector, $a_{i}^{(0)}$, transforms inhomogeneously under Weyl rescaling and together with the boundary metric $\gamma_{ij}^{(0)}$ define a boundary Weyl manifold. The subleading terms of the asymptotic expansion transform as Weyl tensors-- a key aspect for expressing boundary data in terms of Weyl quantities. 

Next, we analyzed Einstein's equations, solving them order by order in the asymptotic expansion, which enables us to relate the expansion coefficients of the boundary metric to those of $a_{i}$. The data needed to holographically reconstruct the spacetime is different from the one in the standard FG case. In the gFG, it is necessary to specify not only the boundary metric $\gamma_{ij}^{(0)}$ but also the full expansion of the vector $a_{i}$.  From these results, we derived the renormalized action, consisting of the standard Balasubramanian--Krauss action supplemented with an additional boundary term  -- a total derivative -- which is necessary to cancel subleading divergences, particularly in topologically non-trivial spacetimes. In that scenario, this total derivate would give rise to a codimension-two term which has been previously denoted in the literature as \textit{corner term} \cite{Adami:2023fbm,Ciambelli:2023ott}. 

The renormalized action enabled us to obtain the holographic stress tensor, which we write in terms of boundary Weyl tensors. Notably, the trace of this stress tensor splits into a Weyl-invariant part and an inhomogeneous component identified as the divergence of a Weyl current sourced by the boundary Weyl connection, recovering the standard Weyl anomaly. 

Finally, we discussed two scenarios where this formalism could play a significant role in obtaining holographic data. First, we briefly examined its application to an integrable model in three dimensions \cite{Cardenas:2021vwo}. In this model, the black hole solutions naturally contain cross terms between the holographic and boundary coordinates, thereby generating a boundary Weyl connection  $a_{i}^{(0)}$ and subleading Weyl tensors $a_{i}^{(n)}$, $n\ge 0$. Interestingly, all solutions contained a non-trivial shift vector, suggesting that breaking the gFG gauge as shown in \autoref{App:pneq1} would be advantageous in this context.

Second, we applied our methods to the accelerating black hole in 2+1 dimensions. We identify a coordinate transformation that renders the metric of an accelerating BTZ black hole into the gFG gauge, introducing a boundary Weyl connection and full asymptotic expansion of the vector $a_{i}$. This transformation introduces free parameters $\zeta$ that do not affect the boundary metric or thermodynamic quantities. However, these coefficients play an important role in the holographic Weyl anomaly as the Weyl current is now directly proportional to it. 
Additionally, the renormalized action is now realized by the Balasubramanian--Krauss action supplemented with the \textit{corner term} \eqref{cod2counter} which regulates the divergences appearing at the intersection between the conformal boundary and the domain wall that is responsible for the acceleration. This new counterterm replaces -- on-shell -- the GHY term of the internal boundary produced by the wall, leading to the same Euclidean on-shell action obtained in \cite{Arenas-Henriquez:2023hur}.

There are several intriguing questions and research directions opened up by this work. We list some of these below:

\paragraph{Other dimensions and flat limit.} A straightforward path to follow is to extend the gFG gauge to higher and lower dimensions.
For instance, while the gFG gauge has already been applied in two-dimensional dilaton gravity \cite{Grumiller:2013swa}, the connection with boundary Weyl structures remains unexplored. Investigating this could lead to interesting holographic implications for the Sachdev--Ye--Kitaev model (see \cite{Maldacena:2016hyu, Cotler:2016fpe, Jensen:2016pah, Engelsoy:2016xyb} and reference therein for the AdS$_2$/CFT$_1$ dictionary).

Additionally, extending the gFG framework to higher dimensions, and in particular four dimensions, is another promising direction. This would allow us to use the techniques developed in this paper to study four-dimensional accelerating black holes. Holographically, such black holes correspond to a strongly/weakly coupled plasma on a black hole background. This might be appropriate for studying relevant non-perturbative quantum states through AdS/CFT \cite{Hubeny:2009kz, Anabalon:2018ydc, Anabalon:2018qfv, Cassani:2021dwa}. 
%
%Recently, it has been shown that accelerating black holes can be embedded into
%In the supersymmetric regime (in which the black hole horizon becomes a spindle), there is no need for extra counterterms of higher codimension and can be uplifted on regular Sasaki--Einstein spaces to give solutions of $D=11$ supergravity \cite{Ferrero:2020twa}. 
%
Analogous to the three-dimensional case presented in \autoref{Sec:Examples}, these solutions feature a cosmic string in their interior extending from the boundary to the black hole horizon \footnote{A more general class of solutions contains two different Killing horizons, the black hole horizon and a cosmological horizon due to the acceleration. Nonetheless, in the AdS case, both horizons can collapse into a single one by tuning the value of the acceleration and the cosmological constant \cite{Dias:2002mi}. In this way, the system achieves thermal equilibrium even though the acceleration persists \cite{Appels:2016uha}. We are considering this scenario when discussing the holographic boundary theory.}. In the dual field theory, this setup corresponds to modeling an external quark with computable drag force and mass related to the string tension and boundary conditions \cite{Gubser:2006bz}. For finite mass, due to Hawking radiation, the string becomes excited generating transversal fluctuations and resulting in a Brownian motion at the string's endpoint \cite{deBoer:2008gu, Atmaja:2010uu}. Such behaviour has not been studied in the theory dual to accelerating black holes although the geometrical construction of these solutions naturally inherits the setup to study holographic Brownian motion. It would be interesting to see if employing the gFG gauge sheds light on this regard.

Another interesting direction is to investigate the most general set of boundary conditions within the framework of flat holography. A flat-space analogue of the gFG gauge was introduced in \cite{Grumiller:2017sjh}. By adopting a similar line element to \eqref{gFG} (without factors of $\ell$) and imposing Ricci flatness as on-shell constraint, one obtains a geometry that encompasses various previously studied boundary conditions, such as Barnich--Compere/ $\mathfrak{bms}_3$ \cite{Barnich:2006av}, Heisenberg \cite{Afshar:2016wfy}, and Detournay--Riegler \cite{Detournay:2016sfv}. 
Consistently, the same set of on-shell constraints also emerge in the CS formulation. 
The extension of the gauge/gravity duality to asymptotically flat spacetimes has become an important area of research, offering potential insights into quantum gravity in an asymptotically flat universe. It is conjectured that the dual field theory at the null boundary of such spacetimes corresponds to a BMS/Carrollian Conformal Field Theory \cite{Barnich:2010eb, Barnich:2013yka, Duval:2014uva, Hartong:2015usd,Bagchi:2016bcd,Donnay:2022aba,Donnay:2022wvx}. More recently, Carrollian structures have proven essential in understanding diverse non-relativistic phenomena such as quantum anomalies, hydrodynamics, and strongly correlated systems, see for instance \cite{Marsot:2022imf,Figueroa-OFarrill:2023vbj,Perez:2023uwt,deBoer:2017ing,deBoer:2021jej,Kolekar:2024cfg} and reference therein.
A holographic analysis within this general gauge framework could open up new avenues for exploring boundary dynamics and asymptotic charges, particularly in connection with Weyl structures, which constrain the dual theory by enforcing Weyl covariance. One notable application involves Carrollian fluids \cite{Ciambelli:2018wre,Ciambelli:2018xat,Petkou:2022bmz, Athanasiou:2024ykt,Adami:2024rkr}, where the coupling of Weyl invariance with Carrollian symmetries could provide a deeper understanding of fluid dynamics in ultra-relativistic or non-relativistic regimes. Such an approach could, for instance, elucidate the role of conformal Carroll symmetry in holography and its relationship with non-relativistic gravitational theories, offering fresh perspectives on scaling limits, boundary quantum states, and holographic transport phenomena.
Significant progress in that direction has already been made in the context of three-dimensional gravity, where the asymptotic symmetries of the gravitational solutions are mapped to the symmetries of dual Carrollian fluids  \cite{Ciambelli:2020ftk,Ciambelli:2020eba,Campoleoni:2022wmf}.

An interesting feature of the CS formulation of \cite{Grumiller:2016pqb} is its ability to consistently go from AdS$_3$ to flat space via a \.{I}n\"{o}n\"{u}--Wigner contraction of the AdS generators, resulting in the global Carroll algebra. In fact, this contraction corresponds to the ultra-relativistic and flat limit taken simultaneously leading to a new Carrollian gravitational theory.  In the metric formulation, such a limit cannot be taken directly in the gFG gauge as \eqref{gFG} contains positive powers of the AdS radius. Therefore, a different gauge must be used to obtain flat holography from a limit of the AdS/CFT correspondence. Important progress has been made in this direction, see for instance \cite{Campoleoni:2023fug, Bagchi:2023fbj,Alday:2024yyj,Kraus:2024gso}. It would be interesting to use a gauge which admits a boundary Weyl structure to describe asymptotically AdS space with a well-defined flat limit to obtain information on the dual theory of flat spacetime from the AdS/CFT correspondence.

\paragraph{Entanglement entropy.}

Another interesting direction to explore is how the relaxation of the gauge affects the boundary quantum states. A simple setup that could shed light on this effect is the notion of entanglement in the dual theory. Using the Ryu--Takayangi (RT) prescription \cite{Ryu:2006bv}, where the entanglement entropy of a region $A$ in the boundary CFT is proportional to the area of a minimal spacelike surface $\gamma_A$ (referred to as the RT surface) that ends on the boundary of $A$ in the bulk AdS space\footnote{See \cite{Hubeny:2007xt} for the covariant generalization.}, we may gain new insights into how the gauge structure influences entanglement properties in the holographic dual. In three dimensions, we consider the holographic coordinate to be parametrized by the spatial coordinate, i.e., $z = z(x)$, and considering a system on a region of length $l$, i.e., $-\nicefrac{l}{2} < x < \nicefrac{l}{2}$~, subject to the boundary condition that the RT surface is anchored to the entangling region, i.e., $z(\pm l/2) = 0$~, the entanglement entropy in the gFG gauge \eqref{gFG} reads
\begin{align}\label{SRT}
    S_{\rm E} = \frac{{\rm A}_{\rm RT}}{4G} = \frac{\ell}{4G}\int_{-l/2}^{l/2} dx \sqrt{\left(\frac{z'}{z}\right)^2+\frac{h_{xx}}{\ell^2} - a_x\left(\frac{z'}{z}\right) +  a_x^2 }~.
\end{align}
Turning off the vector component $a_i = 0$ recovers the standard AdS$_3$/CFT$_2$ result to obtain the entanglement entropy of a strip of length $l$ when considering the Poincar\'e patch of AdS$_3$. 
Minimizing the RT surface in the presence of $a_i$ becomes a highly non-trivial task. The resulting Euler--Lagrange equation associated with \eqref{SRT} is a second-order non-linear differential equation that is hard to solve even in simplified setups.
An analytical solution can be obtained in the case of a flat boundary $h_{ij} = \eta_{ij}$ and the case $a_i=\alpha/z$ with $\alpha\in \mathbb{R}$~. However, in this case, we find that no solution satisfies the RT boundary conditions unless $\alpha=1$~. For this particular value, the bulk metric becomes singular. An additional solution can be found for $\alpha=2$ where the entanglement entropy vanishes. Therefore, this highlights that computing entanglement entropy in AdS$_3$/CFT$_2$ depends heavily on the chosen coordinates. It would be interesting to explore how deviations from the FG gauge affect the holographic entanglement entropy by finding exact solutions to the RT problem in the gFG gauge. 

Moreover, as we have shown in previous sections, the gFG gauge is suitable for studying holographically accelerated black holes. In \cite{Tian:2023ine,Tian:2024mew} is suggested that the holographic entanglement is modified by the presence of the cosmic brane generating a boundary entropy \cite{Takayanagi:2011zk}. The topological defect generates a boundary in the dual CFT, and therefore, there should be an extra contribution to the entanglement entropy coming from such a boundary. It would be interesting to see how the new counterterm \eqref{cod2counter} (that is not trivial for the accelerating BTZ black hole) can be used to obtain such a contribution.

Furthermore, as shown in \cite{Naseh:2016maw}, a topological QFT with scale invariance would possess a Weyl current that modifies the entanglement entropy. 
In this case, quantum states are mixed with classical correlations for mixed states. Therefore, it would be interesting to study logarithmic negativity \cite{Kudler-Flam:2018qjo,Kusuki:2019zsp} and other quantum measures that could allow us to understand points of an RG flows in the holographic setup and adapt them for the gFG and WFG gauges. 
As discussed in \cite{Ciambelli:2023ott}, the transformation of the Weyl source is now associated with a one-form symmetry similar to a Wilson line in the dual theory. A Wilson line encircling the entangling region modifies the holographic entanglement entropy \cite{Belin:2013uta}. It would be intriguing to construct a similar modification of the entanglement entropy produced by a Weyl--Wilson line and understand it from the dual field theory perspective. 

\paragraph{Weyl charges.}
As explained in \autoref{Sec:Intro}, Penrose conformal compactification of AdS$_3$ spacetimes induces a conformal class of boundary metrics $[\gamma^{(0)}]$. Nevertheless, it is still possible to unravel the properties of asymptotically AdS spaces without fixing the boundary metric explicitly. In particular, when considering the asymptotic symmetries associated with AdS$_{3}$ in the FG gauge, contain the usual two copies of the diffeomorphisms of $S^1$ and an additional Weyl symmetry due to the freedom of considering Weyl rescalings of the boundary metric $\gamma^{{(0)}}$~ \cite{Alessio:2020ioh}. The associated Weyl surface charges are integrable but in general, not conserved due to their time dependence, see for instance \cite{Barnich:2007bf,Adami:2020ugu} for discussions on this matter.  At the quantum level, this is translated into an anomalous Ward--Takahashi identity associated with the Weyl rescaling of the holographic theory. The asymptotic Killing vectors are enhanced in comparison with the ones of Brown and Henneaux by considering a Weyl rescaling of the boundary metric and allowing the Weyl factor $\omega$ to fluctuate inducing a new radial component $\xi^z = z \omega(x^i)+\mathcal{O}(z^{2})$~. The same asymptotic Killing vectors were obtained in the WFG gauge in \cite{Ciambelli:2023ott}, showing that both gauges share the same family of asymptotic algebras without inducing any Weyl transformation on the boundary fields, as also confirmed in the Chern--Simons formulation. 

The associated asymptotic symmetry algebra corresponds to the standard two sums of centrally extended Virasoro algebra with an extra centrally extended Weyl sector, i.e.,
\begin{align} \label{asympsymm}
    \{Q_{\xi^\pm_n},Q_{\xi^\pm_m}\} ={}& i(n-m)Q_{\xi^\pm_{n+m}} - im^3\frac{c^\pm}{12}\delta_{n+m,0}~, \nonumber \\ \{Q_{\zeta_{pq}},Q_{\zeta_{rs}}\} ={}& -i(r-q)c_W \delta_{p+r,q+s}~,
\end{align}
with no other non-trivial commutator. 
The central extension of the asymptotic symmetry algebra exhibits a field-dependent Lie algebra 2-cocycle, a characteristic feature of odd-dimensional AdS gravity (even dimensional boundary) described in the FG gauge \cite{Fiorucci:2020xto} due to the Weyl anomaly.  Notably, a similar field-dependent Lie algebra 2-cocycle was first observed in the gravitational charge algebra of four-dimensional asymptotically flat spacetimes \cite{Barnich:2011mi}.

The Virasoro central charges are the ones obtained by Brown and Henneaux $c^\pm =\nicefrac{3\ell}{2G}$~, and the Weyl central charge
\begin{align}
    c_W = \frac{\ell}{2G}\exp\{2i(q+s)t/\ell\}
\end{align}
is explicitly time-dependent. Therefore, there is a family of algebras on each given time slice parametrized by mode numbers $\zeta_{pq}$~. In the case that the conformal factor admits a chiral splitting, i.e. it becomes box-free in light-cone coordinates, the Weyl central charge becomes constant and time-independent showing that the asymptotic symmetry algebra is a direct sum of two copies of a Virasoro algebra and a $U(1)$ current algebra \cite{Troessaert:2013fma}.

An intriguing aspect is that the associated Weyl charges do not manifest in the metric version of CS gravity \cite{Ciambelli:2024vhy}. Furthermore, in Einstein gravity, these charges vanish when employing the Bondi--Sachs gauge -- a gauge well-suited for directly taking the flat limit \cite{Compere:2019bua,Compere:2020lrt} -- unlike the FG gauge. Interestingly, the opposite behaviour occurs in the metric formulation of CS gravity, where the Weyl charges remain present.

In the gFG gauge, the asymptotic charges and their associated algebra may get a modification due to the new decay of the bulk metric, as shown in \cite{Grumiller:2016pqb}. This would imply that there might be new boundary sources. For instance, we could consider variations with respect to other components of the vector expansion, say $a_i^{(-1)}$~, and find a new current proportional to $a_i^{(1)}$~. In \cite{Grumiller:2016pqb}, the asymptotic symmetry algebra associated with the gFG gauge is studied in detail in the Chern--Simons formalism of three-dimensional AdS gravity, finding two copies of affine ${\mathfrak sl}(2)_k$-algebras. Nonetheless, when considering the gFG gauge in the metric formulation the authors find that the algebra of asymptotic Killing vectors does not close under the ordinary Lie bracket and one needs to consider an extended version of the bracket \cite{Barnich:2010eb, Compere:2015knw} and find an affine current algebra but without central extension, in contrast with the Chern--Simons formalism. Furthermore, the algebra of asymptotic charges was not computed in the metric formalism but rather an algebra of asymptotic Killing vector fields. This shows some subtle differences when translating between the Chern--Simons and metric formalism. Moreover, in \cite{Grumiller:2017sjh} certain components of the metric and vector expansion are allowed to vary while others are held fixed at the boundary, simplifying the discussion and analysis of the asymptotic algebra. Relaxing this condition, as shown in \cite{Perez:2016vqo}, allowing variations in all components of the metric leads to an infinite set of Hamiltonian generators associated with the KdV hierarchy. 

A powerful approach for extracting asymptotic charges involves utilizing the covariant phase space formalism to compute the renormalized symplectic potential. This method is particularly sensitive to corner terms \cite{McNees:2023tus}, allowing modifications to boundary sources and, consequently, offering a new holographic interpretation of the symplectic potential. In our future work, we aim to explore the resulting asymptotic symmetry algebra and its holographic implications. Specifically, we will investigate the implications and dual interpretation of the corner term \eqref{cod2counter}, extending the framework of \cite{Ciambelli:2023ott} within the gFG gauge.

Along these lines, a non-trivial physical charge arises when transitioning from the FG gauge to the WFG gauge in three dimensions \cite{Ciambelli:2023ott}. Then, fixing the gauge corresponds to making use of a charged bulk diffeomorphism that changes the degrees of freedom of the holographic theory. This was noticed by performing an asymptotic 
analysis of the fields finding that there is a new charge associated with the leading order of the vector expansion $a_i^{(0)}$~. It would be interesting to see whether the transformation that brings the gFG gauge to the WFG gauge is charged. We leave this question open for future work.

\section*{Acknowledgements}

We would like to express our gratitude to Jos\'e Barrientos, Kedar Kolekar, Rodrigo Olea, Leonardo Santilli, Wei Song, Per Sundell, and Shing-Tung Yau, for comments and discussions on related topics. Specially, we would like to thank Hamed Adami and Luca Ciambelli for carefully reading the draft and providing us with valuable comments and suggestions for improving the quality of the paper. The work of
{\sc Gah} is supported by the Beijing Natural Science Foundation (IS24016) and the Shuimu Scholars program of Tsinghua University. 
The work of FD and DRB was supported by Anillo Grant ACT210100 \textit{Holography and its applications to High Energy Physics, Quantum Gravity and Condensed Matter
Systems}.

\begin{appendix}
\section{Generalized AdS$_3$ boundary conditions}\label{App:pneq1}
Consider the most general AlAdS geometry proposed in \cite{Grumiller:2016pqb},
\begin{align}\label{gFGpneq1}
    ds^2 = \ell^2 {\cal F}(z,\vec{x})^2\left(\frac{dz}{z}-{ a}_i(z,\vec{x}) dx^i\right)^2 + h_{ij}(z,\vec{x})dx^idx^j~,
\end{align}
which leads to the gFG gauge \eqref{gFG} when fixing the lapse function ${\cal F}^2=1$~. To ensure the solution asymptotically approaches AdS$_3$, we impose the asymptotic behaviour \eqref{asybeh} and
\begin{align}
    {\cal F} = \sum_{k\geq 0}\left(\frac{z}{\ell}\right)^{k}{\cal F}_{(k)}(x^i)~,
\end{align}
so that the boundary metric becomes
\begin{align}
    \gamma^{(0)}_{ij}=h_{ij}^{(0)} + \ell^2 {\cal F}_{(0)}^2 a_i^{(-1)}a_j^{(-1)}~.
\end{align}
Under the bulk diffeomorphisms \eqref{WeylDiff}, the metric transforms as 
\begin{align}
    ds'{}^2 = \ell^2 {\cal F}(\cB z',x')^2\left(\frac{dz'}{z'}-a'_i(\cB z',x')dx'{}^i\right)^2 + h_{ij}(\cB z',x')dx'{}^i dx'{}^j~,
\end{align}
where we use the transformation rule \eqref{aprime} and 
\begin{align}
    {\cal F}(\cB',x') = {\cal F}_{(0)}(x') + \frac{z'}{\ell}\cB {\cal F}_{(1)}(x') + \left(\frac{z'}{\ell}\right)^2 \cB^2 {\cal F}_{(2)}(x') + \dots~.
\end{align}
Thus, the expansion coefficients of the metric function ${\cal F}(z,x^i)$, transform Weyl covariantly with definite Weyl weights $\cB^{k}$, i.e.,
\begin{align}
    {\cal F}_{(k)}\to \cB^k {\cal F}_{(k)}~,\qquad \forall k\geq 0~,
\end{align}
and that the general line element structure \eqref{gFGpneq1} remains stable under \eqref{WeylDiff} still inducing a Weyl connection at the boundary.

\paragraph{Field Equations.}  In this section, we will consider covariance with respect to $h_{(0)}$ rather than $\gamma_{(0)}$. In this case, the analysis of the accelerating black hole in \autoref{Sec:Examples} simplifies notably. Nonetheless, going back to the original boundary metric $\gamma_{(0)}$ is nothing else than a simple exercise. Repeating the process, we solve Einstein's equations order by order in $z$, finding the leading order
\begin{align}
    a_i^{(-1)}a_j^{(-1)}h_{(0)}^{ij} = \frac{1}{\ell^2}\left(1-\frac{1}{{\cal F}_{(0)}^2}\right)~. 
\end{align}
For the metric coefficients, we get
\begin{align}
    h_{ij}^{(1)} ={}& \alpha_1 \hat{\cD}_{(i} a^{(-1)}_{j)} + \alpha_2 \pone a^{(-1)}_i a^{(-1)}_j +\alpha_3 \pone h_{ij}^{(0)}a^{(-1)}_m a^m_{(-1)}\nonumber\\
    {}&+\alpha_4 a^{(-1)}_{(i}\hat{\cD}_{j)}\pzero+\alpha_5 h_{ij}^{(0)}a^m_{(-1)} \hat{\cD}_m \pzero+\alpha_6a^{(-1)}_{(i}\ast\!a^{(-1)}_{j)}\hat{\cD}_m\ast\!a^{m}_{(-1)}\,,
\end{align}
where 
\begin{align}
    \alpha_1 = -2\ell^2~,\quad \alpha_2 = -\frac{4\ell^2}{\pzero}~,\quad \alpha_3 = \frac{2\ell^2(\pzero^2-2)}{\pzero(\pzero^2-1)}~,\nonumber\end{align}\vspace{-0.5cm}\begin{align} \alpha_4 = -\frac{4\ell^2}{\pzero}~,\quad \alpha_5 = \frac{2\ell^2}{\pzero}~,\quad \alpha_6=-2\ell^4\pzero^2\,,
\end{align}
$\hat{\cD}$ is the Weyl covariant derivative with respect to $h_{(0)}$ whose action on Weyl scalars and vectors is
\begin{align}\label{Weyl_derivative}
    \hat{\cD}_iS={}&\nabla_i^{h}S+wa^{(0)}_iS\,,\\
    \hat{\cD}_iV_j={}&\nabla_i^h V_j+(w+1)a^{(0)}_iV_j+a^{(0)}_jV_i-h^{(0)}_{ij}a^m_{(0)}V_m\,,
\end{align}
with $\nabla^h_i$ the covariant derivative respect to $h_{(0)}$~, and a dual vector is defined as
\begin{align}
    \ast V_i := h^{jk}_{(0)}\eta_{ki} V_j~,\qquad \eta_{ij} := \sqrt{-h_{(0)}}\epsilon_{ij}~.
\end{align}
Then,
\begin{align}
    h_{(0)}^{ij}h_{ij}^{(1)} = -2\ell^2\hat\cD_i a^i_{(-1)} - \frac{4\pone}{\pzero^3}~.
\end{align}
Finally, 
\begin{align}
    h^{(2)}_{ij}h^{ij}_{(0)} ={}& \beta_1\hat{R}^h + \left(\beta_2 a^i_{(-1)}\hat{\cD}_i\pzero + \beta_3 \hat{\cD}_i a^i_{(-1)}+\beta_4 \pone\right)\pone + \beta_5h^{ij}_{(0)}\hat{\cD}_i\hat{\cD}_j \pzero\nonumber 
    \\{}& +\beta_6 a^i_{(-1)}\hat{\cD}_i\pzero \hat{\cD}_j a^j_{(-1)} + \beta_7\hat{\cD}_i\pzero \hat{\cD}^i\pzero + \beta_8 \ptwo +\beta_9 a^i_{(-1)}\hat{\cD}_i\pone\nonumber 
    \\ {}& +\beta_{10}\hat{\cD}^i a^j_{(-1)} \hat{\cD}_j a^{(-1)}_i+\beta_{11}\ast\!a^{i}_{(-1)}\hat{\cD}_j\hat{\cD}_{i}\ast\!a^{j}_{(-1)}+\beta_{12}\ast\!a^{i}_{(-1)}\hat{\cD}_{i}\pzero\hat{\cD}_j\ast\!a^{j}_{(-1)}\nonumber
    \\ {}& +\beta_{13}\hat{\cD}_i\ast\!a^{i}_{(-1)}\hat{\cD}_j\ast\!a^{j}_{(-1)}\,,
\end{align}
where $\hat{R}^h$ is the Weyl-Ricci scalar of $h_{(0)}$ and
\begin{align}
  \nonumber  \beta_1 ={}& -\frac{\ell^2}{2}~,\quad \beta_2 = \frac{8\ell^2}{\pzero^2}~,\quad \beta_3 = \frac{4\ell^2(1-\pzero^2)}{\pzero^3}~,\quad \beta_4 = \frac{4\left(1-\pzero^2+\pzero^4\right)}{\pzero^6}~,\\ {}& \beta_5 = -\frac{\ell^2}{\pzero}~,\quad \beta_6 = -\frac{\ell^4\left(3\pzero^2 -2\right)}{\pzero\left(\pzero^2-1\right)}~,\quad \beta_7 = \frac{\ell^2\left(4\pzero^4-\pzero^2-2\right)}{\pzero^4\left(\pzero^2-1\right)}~, \quad \beta_8 = -\frac{2}{\pzero}~,\nonumber \\ {}&\beta_9 = -\frac{2\ell^2}{\pzero}~,\quad \beta_{10} = \ell^4~,\quad \beta_{11} = -\ell^4\pzero^2\,,\quad \beta_{12}=\frac{\ell^4\left(\pzero^4+2\pzero^2-2\right)}{\pzero\left(\pzero^2-1\right)}\,,\nonumber
    \\ {}&\beta_{13}=-\ell^4\pzero^2\left(\pzero^2+1\right)\,.
\end{align}
%%%%%%%%%%%%%%%%%
%%%%%%%%%%%%%%%%%
%%%%%%%%%%%%%%%%%

\paragraph{Renormalized action.} Studying the divergences of the EH action and the boundary terms, a possible renormalized action for the family of metrics \eqref{gFGpneq1} is given by adapting the corner term \eqref{tildebeta} to include the contribution from the ${\cal F}$ function, we find that
\begin{align}\label{RegActionPneq1}
    I_{\rm ren,{\cal F}} = \frac{1}{16\pi G}\int_{\cal M}d^3x\sqrt{-g}\left(R+\frac{2}{\ell^{2}}\right) + \frac{1}{8\pi G}\int_{\partial \cal M}d^2x\sqrt{-\gamma}\left[K - \frac{1}{\ell} -\widebar\nabla_i\left(\ell {\cal F} a^i\right)\right]~,
\end{align}
is, indeed, finite and the new counterterm is once again a total derivative, and hence contributes only when the boundary includes non-trivial topological defects. Nonetheless, such corner term is not covariant and just as in the gFG gauge, gives extra finite contributions that do not match the holographic energy coming from the boundary stress tensor breaking the quantum statistical relation and its thermodynamic interpretation. Therefore, we find that the renormalized action \eqref{Sren} is also a well-defined even when breaking the gauge and includes the ${\cal F}$ function. The only difference is that now the boundary component of the normal vector becomes
\begin{align}
    n^i = \frac{N^i}{N} \equiv -\frac{\ell {\cal F} a^i}{\sqrt{1-\ell^2 a_i a^i}}~.
\end{align}

Expanding the renormalized action, one finds that 
\begin{align}
    X^{(1)} ={}& \pzero h_{(0)}^{ij}h_{ij}^{(1)} +2\pone~, \\ X^{(2)} ={}& \pzero h_{(0)}^{ij}h_{ij}^{(2)} - \frac{ \pzero}{2}\left[h_{(1)}^{ij}h_{ij}^{(1)}-\frac12 \left(h_{(0)}^{ij}h_{ij}^{(1)}\right)^2\right]  + \pone h_{(0)}^{ij}h_{ij}^{(1)}+ 2\ptwo~.
\end{align}

Notice that the gFG gauge is smoothly recovered as the special case ${\cal F}^2=1$ (obtained by taking ${\cal F}_{(0)} = 1$ and ${\cal F}_{(k)} = 0 ~\forall k\geq1$)  in all calculations.

\end{appendix}

\newpage
\bibliographystyle{JHEP}
\bibliography{biblio}
\end{document}